\documentclass[conference]{IEEEtran}
\usepackage{amsmath}
\usepackage[]{graphicx,color}
\usepackage{amsmath,amssymb}
\usepackage{bm}
\usepackage{url}
\usepackage{ascmac}

\newcommand{\lH }{\mathrm{H}}

\IEEEsettopmargin{t}{0.8in}

 \usepackage{algorithm}
 \usepackage{algpseudocode}
\makeatletter
\newcommand{\removelatexerror}{\let\@latex@error\@gobble}
\makeatother
\algnewcommand\algorithmicinput{\textbf{Input:}}
\algnewcommand\INPUT{\item[\algorithmicinput]}
\algnewcommand\algorithmicoutput{\textbf{Output:}}
\algnewcommand\OUTPUT{\item[\algorithmicoutput]}

\usepackage{comment}

\title{Deep Unfolded Multicast Beamforming}
\author{
  \IEEEauthorblockN{Satoshi Takabe\IEEEauthorrefmark{1}\IEEEauthorrefmark{2} 
                and Tadashi Wadayama\IEEEauthorrefmark{1}}%\\
  \IEEEauthorblockA{\IEEEauthorrefmark{1}%
		Nagoya Institute of Technology,
		Gokiso, Nagoya, Aichi 466-8555, Japan, 
 		\{s\_takabe, wadayama\}@nitech.ac.jp} %\\
  \IEEEauthorblockA{\IEEEauthorrefmark{2}%
  		RIKEN Center for Advanced Intelligence Project,
  		Nihonbashi, Chuo-ku, Tokyo 103-0027, Japan
                }
}

\begin{document}
%\ninept
%
\maketitle

\begin{abstract}
Multicast beamforming is a promising technique for multicast communication. Providing an efficient and powerful beamforming design algorithm is a crucial issue because multicast beamforming problems such as a max-min-fair problem are NP-hard in general.
Recently, deep learning-based approaches have been proposed for beamforming design. Although these approaches using deep neural networks exhibit reasonable performance gain compared with conventional optimization-based algorithms,
 their scalability is an emerging problem for large systems in which beamforming design becomes a more demanding task.
In this paper, we propose a novel deep unfolded trainable beamforming design with high scalability and efficiency.
The algorithm is designed by expanding the recursive structure of an existing algorithm based on projections onto convex sets  and embedding a constant number of trainable parameters to the expanded network, which leads to a scalable and stable training process. Numerical results show that the proposed algorithm can accelerate its convergence speed by using unsupervised learning, which is a challenging training process for deep unfolding.

\end{abstract}

%\begin{keywords}
%massive MIMO, overloaded MIMO, detection algorithm,  deep learning 
%\end{keywords}

\section{Introduction}\label{sec_1}

With the development of wireless communication technologies, multicast communication including multicast broadcasting 
has been an attractive research field. In multicast communication, a base station (BS) tries to send identical information to multiple users simultaneously. 
In physical layer networks, \emph{multicast beamforming} is proposed as a multicast communication technique in which the BS with multiple antennas generates a beamformer depending on channel state information (CSI)~\cite{Kari}. Although multicast beamforming is featured as a promising technology for the 5th generation wireless networks~\cite{Aran}, 
providing an efficient and powerful beamforming design algorithm has been a crucial issue in the literature
because the multicast beamforming problem is NP-hard in general.
For example,~\cite{Kari} proposes a semidefinite programming (SDP)-based algorithm for a quality-of-service (QoS) beamforming problem. However, because of the SDP relaxation, there is often a large gap between the performance of a designed beamforming vector and its theoretical upper bound. In addition, because the algorithm uses a SDP solver and samples a large number of candidates of beamforming vectors, it is sometimes inefficient in terms of a computational cost.

Recently, deep learning (DL) has been regarded as a promising approach to beamforming design similar to other research fields in physical-layer wireless communications~\cite{Wang}.
The main problem when we apply  DL techniques to beamforming design is that it is impossible to obtain an optimal solution for supervised data in general, which is totally different from the end-to-end approach for signal detection or channel coding~\cite{Hoy}. 
Several studies tackled this problem from different perspectives.  
For example, \cite{Alk} proposes a DL framework for coordinated beamforming based on hybrid learning using training pilot sequences. 
 The authors of~\cite{Kwon} propose a DL architecture in which users choose two beamforming methods effectively.
 In~\cite{Xia}, a DL framework for efficient beamforming design and improving its performance is proposed using hybrid learning based on a sub-optimal weighted minimum mean squared error algorithm. 
 In addition,~\cite{Xia2} discusses a general ``model-driven'' approach using deep neural networks (DNNs) for beamforming design.
These approaches exhibit reasonable performance with a lower computational cost compared with conventional optimization-based algorithms usually in relatively small systems.
However, these algorithms are based on learning DNNs whose number of trainable parameters grows as the system size such as the number of antennas and/or users increases. 
This means that the scalability of training costs possibly becomes a critical problem if we consider a larger system in which beamforming design is a more difficult task.
 In this sense, it is important to investigate a \emph{scalable} and unsupervised DL technique applicable to beamforming design for a large communication system. 

\emph{Deep unfolding}~\cite{LISTA, DU} is another powerful DL technique especially for signal processing and wireless communication~\cite{Stam}.
Unlike standard DNNs, deep unfolding is based on an existing  iterative algorithm.  
By expanding the recursive structure of the algorithm, a signal-flow graph similar to a feed-forward network is obtained.
Then, we can embed trainable internal parameters that control the convergence speed and possibly fixed point.
In this sense, deep unfolding is a model-driven DL technique in a direct way rather than other DNN-based approaches.
In the supervised learning scenario, these parameters are trained using training data that contain a pair of input and corresponding ideal output.
As an advantage of deep unfolding, we can reduce the number of trainable parameters drastically, which leads to a scalable and stable training process.
Deep unfolding has been applied to various problems such as sparse signal recovery~\cite{LISTA, TISTA, TISTA2},
MIMO signal detection~\cite{He, TPG,TPG2}, decoding of error-correcting codes~\cite{Com2,LDPC}, 
and signature design and signal detection in sparsely code division multiple access~\cite{Yama}.
Results of these works suggest that  deep unfolding is also expected to be an effective and scalable approach 
for multicast beamforming although unsupervised learning of deep unfolding has not been studied extensively.  

The goal of this paper is to propose a novel deep unfolding-based algorithm for multicast beamforming design, namely max-min-fair problems.
As described above, the absence of optimal beamforming vectors prevents us to employ supervised training.
We thus attempt  \emph{unsupervised learning of a deep-unfolded algorithm}, which is a challenging task compared with previous studies based on supervised learning. 
As a base of the proposed algorithm, we borrow the structure of projection onto convex set (POCS) with bounded perturbation  proposed by Fink {\it et al.}~\cite{Fink}. This iterative algorithm searches a candidate of a beamforming vector satisfying a convex feasibility problem with adding perturbation to the candidate, which shows remarkable performance improvement compared with a SDP-based algorithm.  By applying deep unfolding, we will train the internal parameters of the algorithm in an unsupervised manner for fast and better approximation.
 
The outline of the paper is as follows. 
Section~\ref{sec_2} describes a multicast beamforming problem and the POCS algorithm.
In Sec.~\ref{sec_3}, we numerically study deep-unfolded POCS for a convex feasibility problem. 
Section~\ref{sec_4} describes the proposed deep unfolding-based beamforming design and shows numerical results compared with other baseline algorithms.  
Section~\ref{sec_5} is a summary of this paper.

\section{Preliminaries}\label{sec_2}

In this section, we first define multicast beamforming problems and then introduce POCS and POCS-BP for approximating the problems.

%\subsection{Notation}
%A vector $\bm{v} = (v_1, v_2,\ldots, v_n)^T \in \mathbb{C}^n$ is a column vector.
%For a complex-valued function $f: \mathbb{C} \rightarrow \mathbb{C}$,
%$f(\bm{v})$ represents 
%the coordinate-wise application of $f$ to $\bm{v} \in \mathbb{C}^n$ 
%such that 
%$f(\bm{v}) := (f(v_1), f(v_2), \ldots, f(v_n))$.
%The $i$-th element of $f(\bm{v})$, i.e., $f(v_i)$ is also denoted by $(f(\bm{v}))_i$.
%The set of consecutive integers from $1$ to $n$ is denoted by $[n] := \{1,2,\ldots, n\}$.
%The cardinality or size of a finite set $A$ is represented by $|A|$.
%The indicator function $\mathbb{I}[cond]$ takes the value $1$ if the $condition$ is true;
%otherwise it takes the value $0$.
%For a complex vector $\bm{z}$,
%$\bm{z}^\ast$ represents its conjugate. %vector.
%For a matrix $\bm{A}:=(a_{ij})\in\mathbb{C}^{m\times n}$, $\bm{A}^\lH := (a_{ji}^\ast)$ is 
%the Hermitian transpose of $A$.

\subsection{Multicast beamforming}

In this paper, we assume a wireless communication system in which a BS with $N$ antennas sends $ x\in \mathbb{C}$
to a multicast group $\mathcal{K}=\{1,\dots,K\}$ containing $K$ users with a single receive antenna. 
Let $\bm{h}_k, \bm{w}\in\mathbb{C}^N$ be a channel vector for the $k$th user ($k\in\mathcal{K}$) and beamforming vector, respectively.  
Then, the received signal for the $k$th user is given as 
$y_k = \bm{w}^\lH\bm h_k x + n_k$, where $n_k\sim \mathcal{CN}(0,\sigma^2)$ represents a complex additive white Gaussian noise with zero mean and variance $\sigma^2$. In addition, we assume that the BS knows CSI perfectly.

We consider the so-called  max-min-fair (MMF) beamforming problem given as
\begin{equation}
\mbox{Maximize}_{\bm{w}\in\mathbb{C}^N }\mbox{min}_{k\in\mathcal{K}} \frac{|\bm w^\lH \bm h_k|^2}{\|\bm w\|_2^2\sigma^2}. %\nonumber 
\label{eq_mmf1}
\end{equation}
In this problem, one searches a complex beamforming vector to maximize the minimum SNR among $K$ users. 
The solution of the problem is identical to that of the following QoS problem up to scaling~\cite{Kari}:
\begin{align}
\mbox{Minimize}_{\bm{w}\in\mathbb{C}^N }\,& \|\bm w\|_2^2 \nonumber \\
\mbox{subject to }& |\bm{w}^\lH\bm h_k|^2\ge \gamma\, (\forall k \in\mathcal K), \label{eq_be1}
\end{align}
where $\gamma$ is a SNR requirement for all users. 
In the QoS beamforming problem, one searches a beamforming vector to maximize the energy efficiency in the system
under SNR constraints~\footnote{We can extend the following discussions straightforwardly to the case in which the SNR requirement of the $k$th user is given by $\gamma_k$.}.
It is computationally hard to exactly obtain optimal solutions of these problems because they are NP-hard. 

To solve these problems approximately, SDP relaxation is often used~\cite{Kari}.
First, using an identity $\mathrm{tr}(\bm v\bm v^\lH) = \bm v^\lH\bm v$ for $\forall \bm v\in\mathbb{C}^N$, the QoS problem (\ref{eq_be1}) is recast as 
\begin{align}
\mbox{Minimize}_{\bm{X}\in\mathbb{C}^{N\times N} }\,& \mathrm{tr}(\bm X) \nonumber \\
\mbox{subject to }& \mathrm{tr}(\bm X\bm Q_k) \ge \gamma\, (\forall k \in\mathcal K), \nonumber \\
& \bm X = \bm X^\lH,\quad \bm X\succeq 	\bm{0},\nonumber\\
&\mathrm{rank}(\bm X) = 1, \label{eq_be2}
\end{align}
where $\bm Q_k = \bm h_k\bm h_k^\lH$. 
Then, the relation between the solution $\bm{w}^\ast$ of (\ref{eq_be1}) and $\bm X^\ast$ of (\ref{eq_be2}) is given as
$\bm{X}^\ast = \bm{w}^\ast(\bm{w}^\ast)^\lH$. Consequently, we can obtain $\bm{w}^\ast$ using $\bm{X}^\ast$ because $\bm{X}^\ast$ is a rank-1 matrix.
However, the rank-1 constraint of (\ref{eq_be2}) makes the problem intractable.

To approximately solve (\ref{eq_be2}) in polynomial time, we instead solve the following relaxed problem without the rank-1 constraint:
\begin{align}
\mbox{Minimize}_{\bm{X}\in\mathbb{C}^{N\times N} }\,& \mathrm{tr}(\bm X) \nonumber \\
\mbox{subject to }& \mathrm{tr}(\bm X\bm Q_k) \ge \gamma\, (\forall k \in\mathcal K), \nonumber \\
& \bm X = \bm X^\lH,\quad \bm X\succeq 	\bm{0}.\label{eq_be3}
\end{align}
This problem is a SDP problem which can be solved in polynomial time using the interior-point method, for example.
In contrast, obtaining a beamforming vector from the solution $\bm{X}_{\mathrm{SDP}}$ of (\ref{eq_be3}) is not straightforward because 
 $\bm{X}_{\mathrm{SDP}}$ is no longer a rank-1 matrix.
In~\cite{Kari}, some randomization techniques are introduced to generate candidates of a beamforming vector from the approximate solution $\bm{X}_{\mathrm{SDP}}$.
Among the candidates, the vector which satisfies all the QoS constraints and has the lowest power is chosen as an approximate solution.

\subsection{Beamforming design by projections onto convex sets} 

Recently, Fink {\it et al.} proposed a promising approximation algorithm for solving (\ref{eq_be2}) based on POCS for a convex feasibility problem and bounded perturbation resilience~\cite{Fink}. 

To apply POCS to the problem, we consider a convex feasibility problem corresponding to (\ref{eq_be3}), not an optimization problem itself.
First, we define the real Hilbert space $\mathcal{H}:=\bm{X}\in\mathbb{C}^{N\times N}; \bm{X}=\bm{X}^\lH\}$
 of complex Hermitian matrices with the inner product 
\begin{equation}
\langle \bm X,\bm Y\rangle := \mathrm{Re}\{\mathrm{tr}(\bm X\bm Y)\}, 
\end{equation}
and consequent Frobenius norm
\begin{equation}
\|\bm X\| :=\sqrt{\langle \bm X,\bm X\rangle} = \sqrt{\mathrm{tr}(\bm X^2)}. 
\end{equation}

Then, a convex feasibility problem finding a matrix $\bm X\in\mathcal{H}$ satisfying all the constraints of (\ref{eq_be3}) and a power constraint is defined.
%which can be solved in polynomial time.
POCS is a sequential projection method to solve such a convex feasibility problem.
It is given as
\begin{equation}
\bm{X}_{t+1} = T_\ast(\bm X_t) := P_{C_{+}} T_{B_P}^\lambda T_{C_K}^\lambda \dots T_{C_1}^\lambda(\bm{X}_{t}),
\label{eq_pocs}
\end{equation}
where $P_B$ represents a projection operator onto the convex set $B$ and 
$T_{B}^\lambda(\bm{X})$ is an operator given by
\begin{equation}
T_{B}^\lambda (\bm{X}) = \bm{X} +  \lambda (P_B(\bm{X})-\bm X),
\end{equation}
for a real scalar $\lambda$ which controls the convergence speed of POCS.
In addition, $C_{+}$, $B_P$, and $C_k$ ($\forall k\in\mathcal{K}$) are convex sets.
The set $C_{+}$ is a positive semidefinite cone corresponding to the constraint $\bm X\succeq \bm{0}$. 
$B_P$ is the half-space $\{\bm X\in\mathcal{H}; \mathrm{tr}(\bm X) \le P\}$ with a parameter $P(>0)$ representing 
the target power of beamforming vector.
$C_k$ is the half-space for the QoS constraint of the $k$th user, i.e, $ \mathrm{tr}(\bm X\bm Q_k) \ge \gamma$.
It is shown that the sequence $(\bm X_t)_{t\ge 1}$ updated by (\ref{eq_pocs}) converges to a solution of the problem
if $\lambda\in(0,2)$.
It is noted, however, that this POCS is practically hard to obtain a good beamforming vector 
because we need to search the value of $P$ corresponding to a solution of (\ref{eq_be3}).
Moreover, the approximation gap due to the lack of the rank-1 constraint is still inevitable even if we tune the value of $P$ appropriately. 

To compensate the lack of the rank-1 constraint, the authors of~\cite{Fink} 
combine POCS with bounded perturbation (BP).
In the algorithm, which is called POCS-BP in this paper, an ad-hoc perturbation toward a rank-1 matrix is added to  
the matrix $\bm X_t$. 
This perturbation is executed by subtracting all the principal components of $\bm X_t$ except for the largest one from $\bm X_t$.
Then, the update rule of POCS-BP is given as
\begin{equation}
\bm{X}_{t+1} = T_\ast( \bm X_t -\tilde \beta_t \bm{\tilde X}_t ),
\label{eq_pocsbp}
\end{equation}
where $(\tilde \beta_t)_{t\ge 1}$ is a real non-negative bounded sequence satisfying $\sum_{t=1}^\infty \tilde \beta_t <\infty $, 
and 
$\bm{\tilde X}_t $ is all the principal components of $\bm{\tilde X}_t $ except for the largest one given as
\begin{equation}
\bm{\tilde X}_t   = \bm X_t - \lambda_{\mathrm{max}}^t \bm{u}_t\bm{u}_t^\lH ,
\label{eq_pocsbp2}
\end{equation}
with the largest eigenvalue $\lambda_{\mathrm{max}}^t$ and corresponding eigenvector $\bm{u}_t$ of $\bm{X}_t$.
A beamforming vector after $T$ iterations is then obtained as the eigenvector corresponding to the largest eigenvalue of $\bm X_T$.

In~\cite{Fink}, it is shown that POCS is bounded perturbation resilient, i.e, POCS-BP always converges to a solution of 
the convex feasibility problem. Moreover, it is numerically shown that manually tuned POCS-BP outperforms the SDP-based algorithms with randomization techniques.
In addition, they also propose a simple version of POCS-BP by omitting $P_{C_+}$ and $P_{B_P}$, which is given by
\begin{equation}
\bm{X}_{t+1} = T_{C_K}^{\lambda} \dots T_{C_1}^\lambda( \bm X_t -\tilde \beta_t \bm{\tilde X}_t ),
\label{eq_pocsbp3}
\end{equation}
where $\lambda$ and $\{\tilde \beta_t\}$ are parameters of the algorithm.
As an advantage of (\ref{eq_pocsbp3}), it reduces the computational complexity related to eigendecomposition for $T_{C_+}^\lambda(\cdot)$
and parameter tuning of $P$, which practically do not affect to the performance of a beamforming vector.
In the rest of this paper, we use  (\ref{eq_pocsbp3}) as the update rule  of BP-POCS.

\section{Acceleration of POCS by deep unfolding}\label{sec_3}

In this section, we demonstrate a deep-unfolded algorithm based on POCS because POCS itself is a useful algorithm for other wireless communication problems~\cite{po}.
The aim of this section is to verify whether  unsupervised learning of deep unfolding can accelerate the convergence speed of POCS.

As an example, we here consider POCS given by 
\begin{equation}
\bm{X}_{t+1} = T_{B_P}^\lambda T_{C_K}^\lambda \dots T_{C_1}^\lambda(\bm{X}_{t}),
\label{eq_pocs2}
\end{equation}
where the projection $P_{C_+}(\cdot)$ onto a semidefinite cone constraint is omitted from (\ref{eq_pocs}) for simplicity.
This POCS solves the following convex feasibility problem.
\begin{equation}
\mbox{Find } \bm{X} \mbox{ s.t. } \bm{X}\in \mathcal S :=\cap_{k\in\mathcal K} C_k\cap B_P. 
\label{eq_pocs3}
\end{equation}

The architecture of deep-unfolded POCS (DU-POCS) is obtained by embedding iteration-dependent trainable parameters to (\ref{eq_pocs2}). Here, we define DU-POCS as
\begin{equation}
\bm{X}_{t+1} = T_{B_P}^{\lambda_t} T_{C_K}^{\lambda_t} \dots T_{C_1}^{\lambda_t}(\bm{X}_{t}),
\label{eq_pocs2du}
\end{equation}
where $\{\lambda_t\}_{t=1}^T$ is a set of trainable parameters in the algorithm.
Although one can embed independent trainable parameters to each operator $T_{B}(\cdot)$, we use the same parameter $\lambda_t$ in the $t$th iteration to reduce the number of trainable parameter. As a result, DU-POCS has only $T$ trainable parameters,  which is constant to the system size such as $N$ and $K$. %This leads to a fast and stable training process.

The training process of deep-unfolded algorithms is usually executed as supervised learning, i.e., minimizing a loss function of their output and a given true solution.  
On the other hand, in the training process of DU-POCS, such supervised data are unavailable because an arbitrary point in the convex set is a possibly true solution.   
We thus train DU-POCS in an unsupervised manner. 
We use a loss function between a point $\bm X\in \mathcal  H$ and half spaces defined by
\begin{equation}
L(\bm{X};\mathcal S) := \sum_{k\in\mathcal K}\mathrm{ReLU}(\gamma-\langle \bm X, \bm Q_k \rangle) + \mathrm{ReLU}(\mathrm{tr}(\bm X)-P),
\label{eq_pocs3}
\end{equation}
where $\mathrm{ReLU}(x):=\max (x,0)$ and thus $L(\bm{X};\mathcal S)\ge 0$.
This function takes zero iff $\bm{X}\in\mathcal S$ because each term represents the gap between $\bm X$ and a constraint of $\mathcal{S}$.   

In the numerical experiment, we set $N=5$, $K=15$, $\sigma=1.0$, $\gamma=1.0$, and $P=0.5$.
The number of iterations of DU-POCS is set to $T=20$.  The initial values of $\{\lambda_t\}_{t=1}^T$ are set to $1.0$. 
DU-POCS is implemented using PyTorch 1.4~\cite{Pytorch}.
In the training process,  $L(\bm{X}_T;\mathcal S)$ is minimized using the Adam optimizer~\cite{Adam} with learning rate $0.003$.
As a mini-batch training, $1000$ mini batches of size 30 are fed to DU-POCS. These batches contain random channel vectors $\{\bm h_k\}_{k=1}^K$. 

\begin{figure}[t]
   \centering
   \includegraphics[width=0.9\hsize]{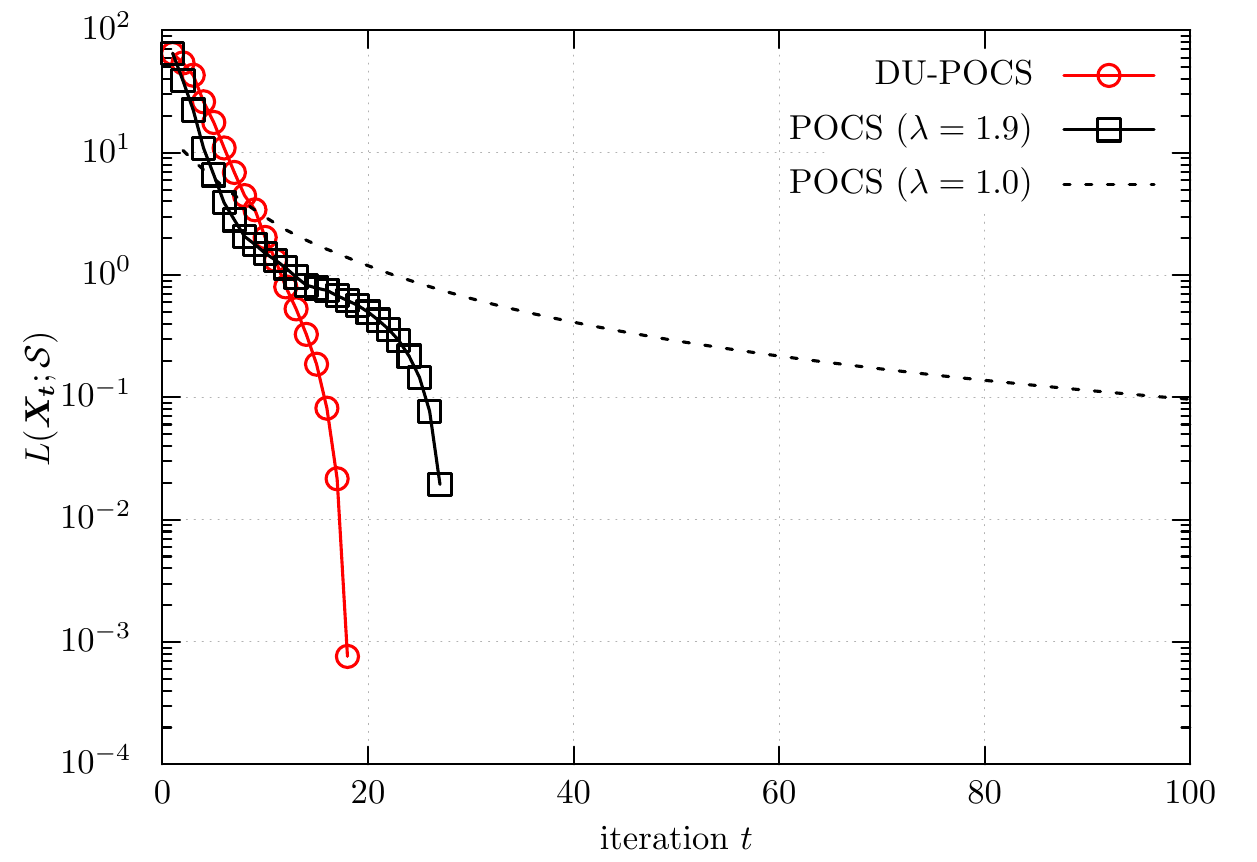}
    \caption{Average loss values of $L(\bm{X}_t;\mathcal{S})$ over 50 realizations as functions of iteration steps $t$. 
    Circles and squares respectively represent DU-POCS and POCS ($\lambda=1.9$). They find a feasible point of a problem within 19 and 28 iterations, respectively. Dotted line represents POCS ($\lambda=1.0$), which does not converge within 5000 iterations.  
    } 
    \label{fig_pocs1}
\end{figure}

Figure~\ref{fig_pocs1} shows the average loss $L(\bm{X}_t;\mathcal S)$ of DU-POCS over 50 realizations.
As a comparison, we show results of POCS (\ref{eq_pocs2}) with $\lambda=1.9$ and $1.0$. 
We find that POCS with $\lambda=1.0$, which corresponds to DU-POCS with initial $\lambda_t$'s, converges very slowly
possibly because a projection onto a half space will break other constraints. It cannot converges to a fixed point within 5000 iterations.
On the other hand, DU-POCS and POCS with $\lambda=1.9$ successfully converges to a point in $\mathcal S$ such that $L(\bm{X}_t;\mathcal S)=0$ within 30 iterations. Namely, DU-POCS converges within 19 iterations whereas POCS with $\lambda=1.9$ does within 28 iterations.
This result suggests that unsupervised learning of DU-POCS successfully accelerates the convergence speed of POCS. 
%by tuning parameters $\{\lambda_t\}$.

\begin{figure}[t]
   \centering
   \includegraphics[width=0.9\hsize]{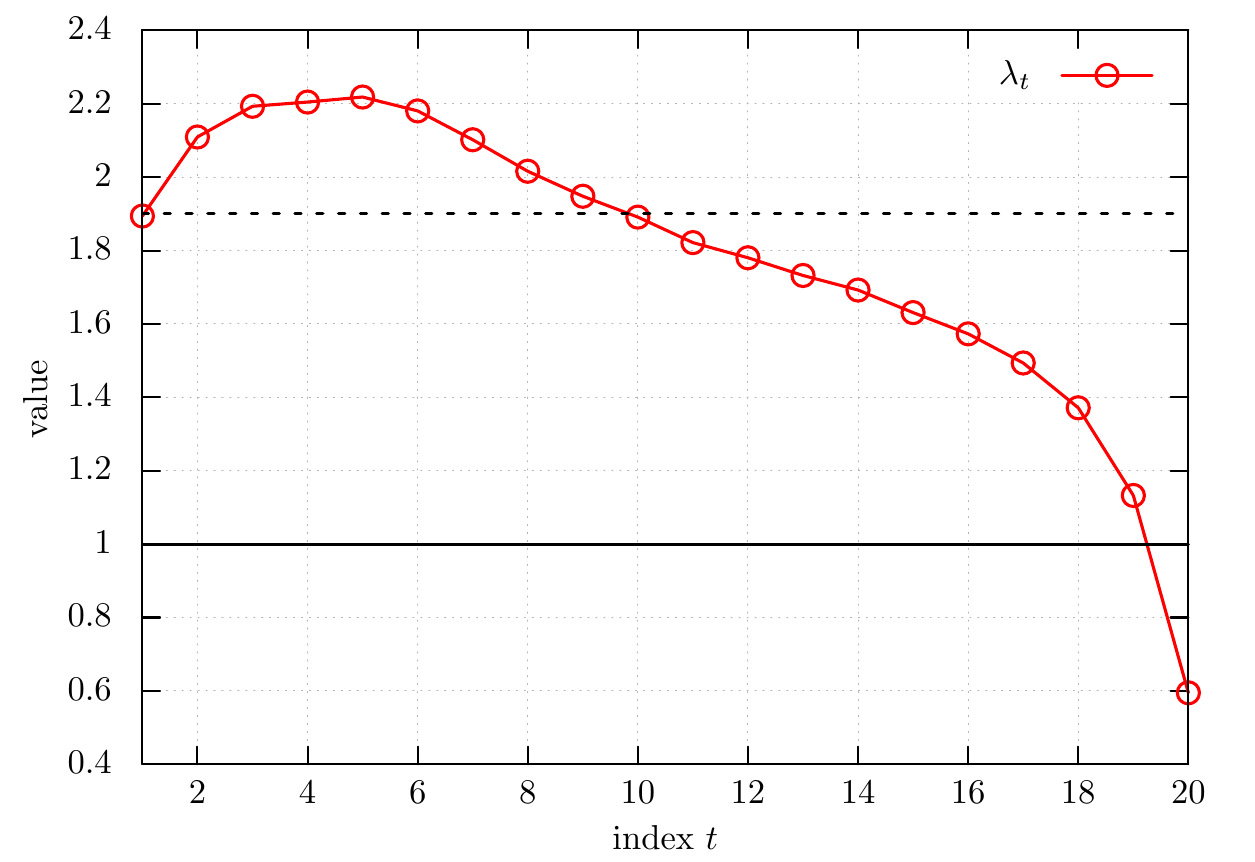}
    \caption{The trained parameters $\{\lambda_t\}_{t=1}^T$ in DU-POCS ($T=20$). The solid line represents $\lambda_t=1.0$ used for the initial values of trainable parameters. The dotted line represents $\lambda_t=1.9$ used in Fig.~\ref{fig_pocs1} for comparison.  
    } 
    \label{fig_pocs2}
\end{figure}

Fig.~\ref{fig_pocs2} shows the trained parameters $\{\lambda_t\}_{t=1}^T$ of DU-POCS. We find that the parameters take different values
depending on the iteration index $t$, which is usually observed in deep-unfolded algorithms.
Recently, the authors theoretically investigate trained parameters of deep unfolded gradient descent and a class of fixed-point iteration algorithms. As a result, it is shown that the so-called Chebyshev steps can reproduce the trained parameters and accelerate the convergence speed~\cite{ch, ch2}.
Although these analyses do not cover POCS (\ref{eq_pocs2du}) in the real Hilbert space, this result suggests that the convergence speed of POCS can also be accelerated by deep unfolding. The theoretical analysis is an interesting future task.

\section{Deep-unfolded POCS-BP}\label{sec_4}
Now we describe the deep-unfolded POCS-BP for multicast beamforming.

From (\ref{eq_pocsbp3}), the update rule of DU-POCS-BP is given by 
\begin{equation}
\bm{X}_{t+1} = T_{C_K}^{\lambda_t} \dots T_{C_1}^{\lambda_t}( \bm X_t -\beta_t^2 \bm{\tilde X}_t ),
\label{eq_dupocs}
\end{equation}
where $\bm{\tilde X}$ is defined as (\ref{eq_pocsbp2}).
The trainable parameters of DU-POCS-BP are $\{\lambda_t,\beta_t\}_{t=1}^T$. Thus, DU-POCS-BP of $T$ iterations has 
$2T$ trainable parameters, which is constant to the system size, $N$ and $K$.
Note that the parameter $\tilde \beta_t$ in POCS-BP is replaced to $\beta_t^2$ to keep coefficients of bounded perturbation non-negative.

The pseudo-code of DU-POCS-BP is shown in Alg. 1. In the algorithm, we need to obtain the maximum eigenvalue and corresponding eigenvector of a matrix. Since Pytorch 1.4 has no function with backward pass for eigendecomposition of a complex matrix,
we alternatively use the power method in Alg. 2 to estimate them. In the following simulations, we use $T_{\mathrm{PM}}=50$ and $\epsilon=10^{-8}$ as parameters of the power method.

\begin{figure}[t]\label{alg}
 \removelatexerror
  \begin{algorithm}[H]
   \caption{DU-POCS-BP}
  \begin{algorithmic}[1]
   \INPUT Channel vectors $\{\bm h_k\}_{k\in\mathcal{K}}$, number of antennas $N$
   \OUTPUT Normalized beamforming vector $\bm w_T\in\mathbb{C}^N$
   \State Initialization: $\bm X =\bm O$
   \For{$t= 1$ to $T$} 
   	\State{Call POWER METHOD to get $\lambda_{\mathrm{max}}$ and $\bm{u}$ of $\bm X$. } 
   	\State{$\bm X := \bm X -\beta_t^2\lambda_{\mathrm{max}} \bm{u}_t\bm{u}^\lH$ }\Comment{Bounded perturbation}
      \For{$k=1$ to $K$}
		\State{$\bm X := T_{C_{k}}^{\lambda_t}(\bm X) $} \Comment{Projection onto $C_k$}
      \EndFor
   \EndFor
   \State{Call POWER METHOD to get $\bm{u}$ of $\bm X$.}
   \State{$\bm w_T := \bm{u}$ }
  \end{algorithmic}
  \end{algorithm}
  
    \begin{algorithm}[H]
   \caption{POWER METHOD}
  \begin{algorithmic}[1]
   \INPUT Diagonalizable matrix $\bm A$
   \OUTPUT Maximum eigenvalue $\lambda_{\mathrm{max}}$ and corresponding normalized eigenvector $\bm u$
   \State Initialization: $\bm u \neq \bm 0$
   \For{$t= 1$ to $T_{\mathrm{PM}}$} 
   	\State{$\bm u' := \bm u$}
   	\State{$\bm u := \bm A\bm u/ \|\bm A\bm u\|_2$}
   	\State{$\lambda_{\mathrm{max}} := \bm u^\lH \bm A\bm u/ \|\bm u\|_2^2$}
   	\If {$\|\bm u-\bm u'\|_\infty<\epsilon$}
   		\State{\textbf{break}}
   	\EndIf
   \EndFor
  \end{algorithmic}
  \end{algorithm}

\end{figure}

\begin{figure}[t]
   \centering
   \includegraphics[width=0.9\hsize]{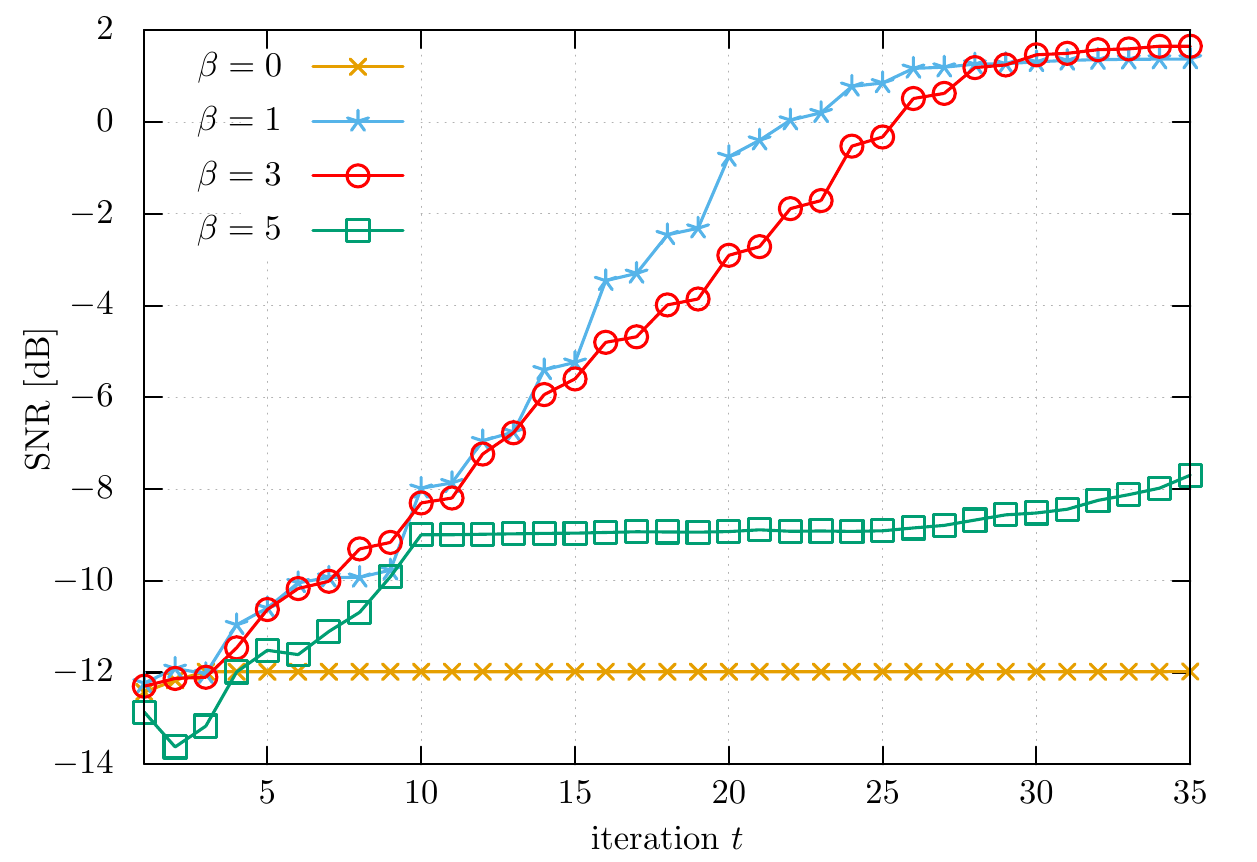}
    \caption{SNR performance of DU-POCS-BP with different values of hyperparameter $\beta$ in the loss function (\ref{eq_dupocs_l1}) when $(N,K)\!=\!(30,50)$.
    SNR is averaged over 50 realizations.
    The hyperparameter $\beta=3$ is used in other experiments.  
    } 
    \label{fig_dupocs3}
\end{figure}

Similar to the case of DU-POCS, it is difficult to obtain an optimal beamforming vector in advance. 
We thus train DU-POCS-BP by unsupervised learning.  
Namely, we try to minimize a target loss function related to the MMF problem (\ref{eq_mmf1}) given by 
\begin{equation}
L(\bm{w}_t) := -\eta\left(\left\{\frac{|\bm w_t\bm h_k|^2}{\sigma^2\|\bm w_t\|_2^2}\right\}_{k\in\mathcal K}; \beta\right),
\label{eq_dupocs_l1}
\end{equation}
where $\bm{w}_t$ is the eigenvector corresponding to the largest eigenvalue of $\bm{X}_t$ and
$\eta(\{s_k\}_{k\in\mathcal{K}};\beta)$ is the weighted softmin function defined by
\begin{equation}
\eta\left(\{s_k\}_{k\in\mathcal{K}}; \beta\right):=\frac{\sum_{k\in\mathcal{K}} s_ke^{-\beta s_k}}{\sum_{k\in\mathcal{K}} e^{-\beta s_k}},
\label{eq_dupocs_l2}
\end{equation}
with the weight $\beta\in\mathbb R$.
Note that the softmin function is used to avoid the differentiable problem of the original min function.
The softmin function is identical to the min function when $\beta\rightarrow\infty$.
In the following experiment, we use $\beta=3$ by tuning it as a hyperparamer (see Fig.~\ref{fig_dupocs3} for details).

Other conditions of the experiments are as follows: we fix $N=30$, $\sigma=1.0$, and  $\gamma=1.0$, and consider two cases in which $K=20$ and $50$.
The number of iterations of DU-POCS-BP is set to $T=35$. The initial values of $\{\beta_t\}_{t=1}^T$ and $\{\lambda_t\}_{t=1}^T$ are respectively set to $0.9^{1/2}=0.948\dots$ and $1.0$.
In mini-batch training, we use incremental training~\cite{TISTA} for stable training. 
In incremental training, randomly generated mini batches are fed to DU-POCS-BP of $T=1$ to train $\beta_1$ and $\lambda_1$. After finishing this, mini batches are fed to DU-POCS-BP of $T=2$ to train $\beta_1$, $\beta_2$, $\lambda_1$, and $\lambda_2$.
In this case, the initial values of $\beta_1$ and $\lambda_1$ are set to trained values in the last mini-batch training.
This procedure is repeated in an incremental manner until $T$ reaches to a given value, 35 in this experiment.     
In each mini-batch training, we used $1000$ mini batches of size $30$. The Adam optimizer with learning rate $0.003$ is executed to minimize the loss function (\ref{eq_dupocs_l1}).

\begin{figure}[t]
   \centering
   \includegraphics[width=0.9\hsize]{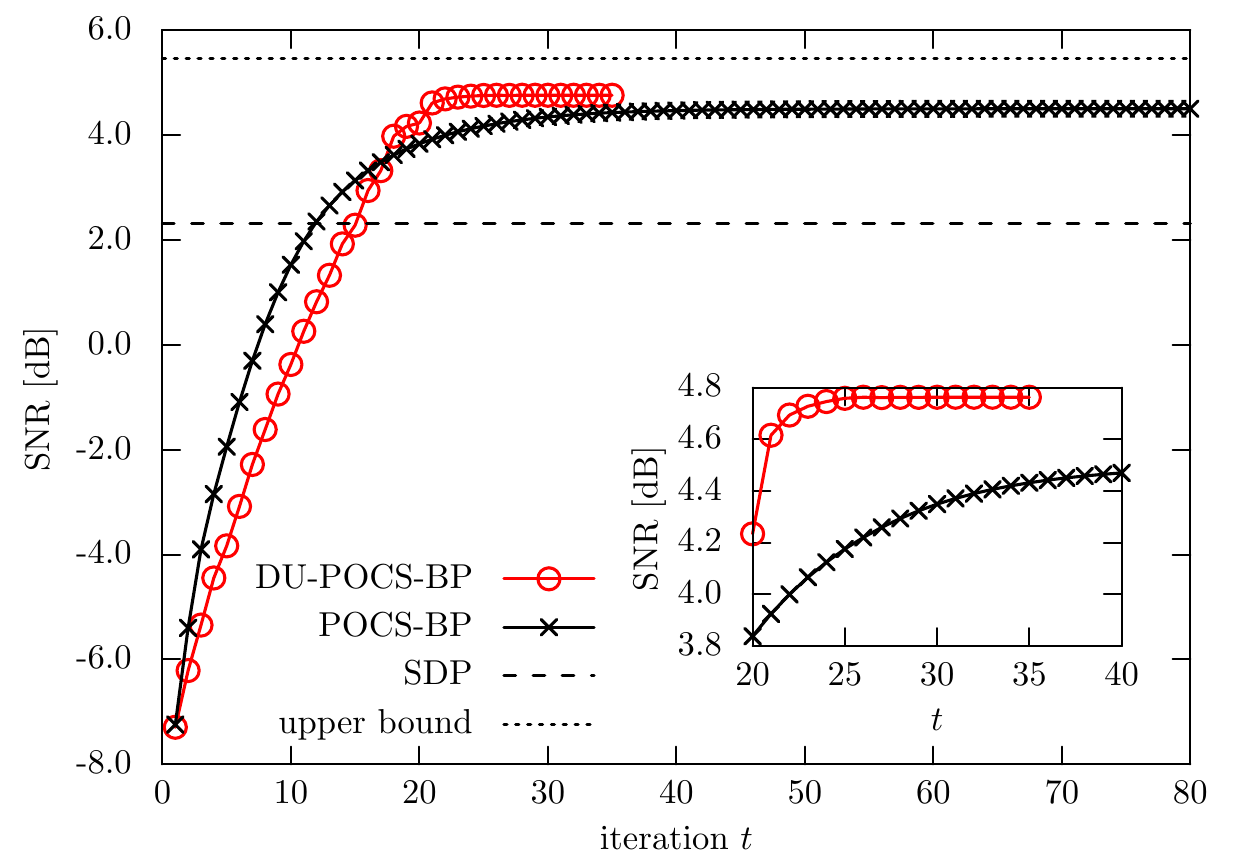}
    \caption{SNR performance of beamforming design algorithms as a function of iteration steps $t$ when $(N,K)=(30,20)$.
    SNR is averaged over 50 realizations.
    Circles and squares respectively represent DU-POCS-BP and POCS-BP.
    Dotted and dashed lines respectively represent the SDP upper bound (\ref{eq_snr2}) and SNR performance of a SDP-based  algorithm.    
    } 
    \label{fig_dupocs1}
\end{figure}

 As a performance measure, we used the minimum SNR among $K$ users given by 
\begin{equation}
\min_{k\in\mathcal{K}} \frac{|\bm w_t\bm h_k|^2}{\sigma^2\|\bm w_t\|_2^2},
\label{eq_snr}
\end{equation}
which is identical to the object function of the MMF problem (\ref{eq_mmf1}). 
The minimum SNR is averaged over 50 realizations of channel vectors.
As a baseline, we execute the original POCS-BP following~\cite{Fink}. In POCS-BP, $\tilde \beta_t=0.9e^{-t/500}$ and $\lambda=1.9$
are used. 
In addition, we also execute the SDP-based beamforming design with randomization in~\cite{Kari}. In the algorithm (called randA in~\cite{Kari}), we first obtain the optimal solution $\bm X_{\mathrm{SDP}}$ of the SDP problem (\ref{eq_be3}). Then, using the eigendecomposition $\bm X_{\mathrm{SDP}}=\bm V\bm \Sigma\bm V^\lH$, we sample candidates of beamforming vectors $\bm{w}=\bm V\bm \Sigma^{1/2} \bm e$, where $\bm e\in\mathbb{C}^N$ is a random vector distributed uniformly on the unit sphere. The sampled $\bm w$ with the largest value of (\ref{eq_snr}) is chosen as the beamforming vector. 
In the experiments, we sample 5000 candidates.
The SDP-based algorithm also provides a SDP upper bound of the MMF problem, which is given by
\begin{equation}
\min_{k\in\mathcal{K}} \frac{\bm h_k^\lH\bm X_{\mathrm{SDP}}\bm h_k}{\mathrm{tr}(\bm X_{\mathrm{SDP}})\sigma^2}.
\label{eq_snr2}
\end{equation}
It is emphasized that this upper bound is not strict in general because the relaxed problem (\ref{eq_be3}) neglects the rank-1 constraint.

Figure~\ref{fig_dupocs1} shows the SNR performance as a function of  the iteration step $t$ when $K=20$. 
 We find that DU-POCS-BP successfully designs a beamforming  vector with smaller number of iterations than POCS-BP.
 Namely,  DU-POCS-BP takes 31 iterations for convergence whereas POCS-BP takes 54 iterations.
As for the convergent SNR, DU-POCS-BP exhibits $4.76$ dB, which is $0.22$ dB larger than POCS-BP as shown in the inset of   Fig.~\ref{fig_dupocs1}. 
Although the gain is relatively small, DU-POCS-BP can converge faster than the original POCS-BP without knowing optimal beamforming vectors in training process.
The SNR performance of these algorithms is higher than that of  the SDP-based algorithm and relatively close to the SDP upper bound.

\begin{figure}[t]
   \centering
   \includegraphics[width=0.9\hsize]{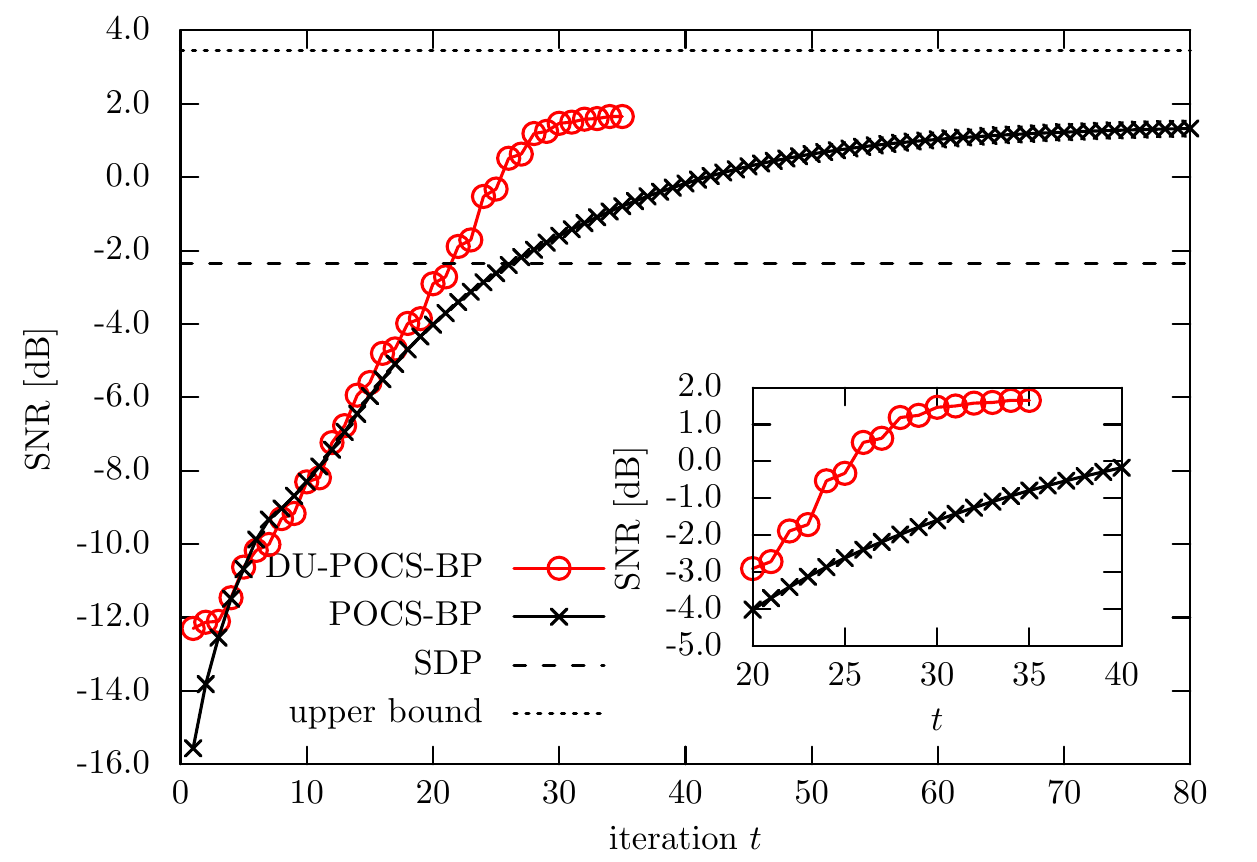}
    \caption{SNR performance of beamforming design algorithms as a function of iteration steps $t$ when $(N,K)=(30,50)$.
    SNR is averaged over 50 realizations.
    Circles and squares respectively represent DU-POCS-BP and POCS-BP.
    Dotted and dashed lines respectively represent the SDP upper bound (\ref{eq_snr2}) and SNR performance of a SDP-based  algorithm.   
    } 
    \label{fig_dupocs2}
\end{figure}

 Figure~\ref{fig_dupocs2} shows the SNR performance when $K=50$, which is more demanding setting for multicast beamforming.
 Similar to Fig.~\ref{fig_dupocs1}, DU-POCS-BP rapidly converges to a fixed point; 
 DU-POCS-BP converges within  34 iterations whereas POCS-BP takes 139 iterations for convergence.
The SNR gain of  DU-POCS-BP is about $0.11$ dB compared with BOCS-BP and about $3.96$ dB compared with SDP.
The results suggest that DU-POCS-BP is effective when the number of users becomes large. 
%where the number of constraints increases.

Figure~\ref{fig_dupocs3} shows SNR performance of DU-POCS-BP with different values of hyperparameter $\beta$ for $\eta(\cdot;\beta)$ in the loss function (\ref{eq_dupocs_l1}).
The resulting SNR  performance largely depends on the value of $\beta$ because $\beta$ decides the ratio between the minimum SNR and other SNRs among users.
In particular, when $\beta=0$, the function $\eta(\cdot;\beta)$ becomes an average, which results in a poor training result.
%In contrast, when $\beta=5$, learning DU-POCS-BP seems difficult  
From this observation, it is crucial to check the $\beta$-dependency in advance, and use the proper value to maximize the convergent SNR performance.

\section{Concluding remarks}\label{sec_5}
In this paper, we propose a deep unfolded multicast beamforming design based on POCS-BP, which is trained by unsupervised learning.
We first demonstrate that deep unfolding can accelerate the convergence speed of POCS without knowing an optimal solution.
Then, we propose a  deep unfolded multicast beamforming design called DU-POCS-BP.
Because DU-POCS-BP has only $2T$ trainable parameters in $T$ iterations, its training process is stable and highly scalable with respect to the system size, $N$ and $K$.
Numerical results show that DU-POCS-BP exhibits acceleration of the convergence speed and small SNR gain compared with the original POCS-BP, which suggests that  DU-POCS-BP is an efficient DL-based beamforming design.
It is a future task to analyze the behavior of DU-POCS and DU-POCS-BP theoretically and extend the proposed method to 
general multicast beamforming design such as multi-group multicast beamforming. 
It would also be an interesting work to demonstrate other applications of unsupervised learning of deep unfolding in wireless communication.

\section*{Acknowledgement}
This work was partly supported by JSPS Grant-in-Aid for Scientific Research (B) 
Grant Number 19H02138 (TW) and Grant-in-Aid 
for Early-Career Scientists Grant Number 19K14613, 
and the Telecommunications Advancement Foundation (ST).

\vfill\pagebreak

% References should be produced using the bibtex program from suitable
% BiBTeX files (here: strings, refs, manuals). The IEEEbib.bst bibliography
% style file from IEEE produces unsorted bibliography list.
% -------------------------------------------------------------------------
%\bibliographystyle{IEEEbib}
%\bibliography{strings,refs}

\end{document}